\documentclass[10pt,conference]{IEEEtran}

\usepackage{amsmath,amssymb,amsfonts}
\usepackage{algorithmic}
\usepackage[linesnumbered,ruled,vlined, longend]{algorithm2e}
\usepackage{svg}
\usepackage{array}
\usepackage{graphicx}
\usepackage{booktabs}
\usepackage{makecell}
\usepackage{comment}
\usepackage{float}
\usepackage{newtxtext}
\usepackage{hyperref}
\usepackage{multirow,soul}
\def\BibTeX{{\rm B\kern-.05em{\sc i\kern-.025em b}\kern-.08em
    T\kern-.1667em\lower.7ex\hbox{E}\kern-.125emX}}

\begin{document}

\title{Q-Cluster: Quantum Error Mitigation Through Noise-Aware Unsupervised Learning}

\author{
\IEEEauthorblockN{Hrushikesh Pramod Patil}
\IEEEauthorblockA{\textit{NC State University} \\
Raleigh, NC, USA \\
hpatil2@ncsu.edu}
\and
\IEEEauthorblockN{Dror Baron}
\IEEEauthorblockA{\textit{NC State University} \\
Raleigh, NC, USA \\
dzbaron@ncsu.edu}
\and
\IEEEauthorblockN{Huiyang Zhou}
\IEEEauthorblockA{\textit{NC State University} \\
Raleigh, NC, USA \\
hzhou@ncsu.edu}

}
\maketitle
\begin{abstract}
Quantum error mitigation (QEM) is critical in reducing
the impact of noise in the pre-fault-tolerant era, and is expected to complement error correction in fault-tolerant quantum computing (FTQC). In this paper, we propose a novel QEM approach, Q-Cluster, that uses unsupervised learning (clustering) to reshape the measured bit-string distribution. Our approach starts with a simplified bit-flip noise model. It first performs clustering on noisy measurement results, i.e., bit-strings, based on the Hamming distance. The centroid of each cluster is calculated using a qubit-wise majority vote. Next, the noisy distribution is adjusted with the clustering outcomes and the bit-flip error rates using Bayesian inference. Our simulation results show that Q-Cluster can mitigate high noise rates (up to 40\% per qubit) with the simple bit-flip noise model. However, real quantum computers do not fit such a simple noise model. To address the problem, we (a) apply Pauli twirling to tailor the complex noise channels to Pauli errors, and (b) employ a machine learning model, ExtraTrees regressor, to estimate an effective bit-flip error rate using a feature vector consisting of machine calibration data (gate \& measurement error rates), circuit features (number of qubits, numbers of different types of gates, etc.) and the shape of the noisy distribution (entropy). Our experimental results show that our proposed Q-Cluster scheme improves the fidelity by a factor of 1.46x, on average,
compared to the unmitigated output distribution, for a set of low-entropy benchmarks on five different IBM quantum machines. Our approach outperforms the state-of-art QEM approaches M3~\cite{m3}, Hammer~\cite{Hammer}, and QBEEP \cite{qbeep} by 1.29x, 1.47x, and 2.65x, respectively.

\end{abstract}


\maketitle

\section{Introduction}
Quantum hardware manufacturers have started demonstrating early quantum error correction on various quantum processors: Google on superconducting quantum processing units (QPUs) \cite{Acharya2023}, Microsoft and Quantinuum on ion trap QPUs \cite{dasilva2024demonstrationlogicalqubitsrepeated}, and Lukin et al. and QuEra on neutral atom QPUs \cite{Bluvstein2024}. There is growing optimism that the era of fault-tolerant quantum computers 
(FTQC) is approaching. However, according to IBM’s roadmap \cite{IBMroadmap}, a fully fault-tolerant quantum computer with more than 1,000 logical qubits remains at least a decade away.

Meanwhile, quantum error mitigation (QEM) has proven to be the most effective way to achieve practical quantum advantage, as demonstrated by IBM \cite{Kim2023}. QEM reduces the effects of quantum errors in circuits by using classical post-processing on an ensemble of circuit runs \cite{RevModPhys.95.045005}. Popular QEM techniques to mitigate expectation values include Zero Noise Extrapolation (ZNE) \cite{ZNE}, Probabilistic Error Cancellation (PEC) \cite{PEC}, and Clifford Data Regression (CDR) \cite{CDR}.
To refine output probability distributions, methods such as Matrix-free Measurement error Mitigation (M3) \cite{m3} and its variant~\cite{PhysRevA.106.012423}, HAMMER \cite{Hammer}, and QBEEP \cite{qbeep} have been proposed. Jigsaw \cite{Jigsaw} also fine-tunes the output probability distribution by mitigating measurement crosstalk errors with subsetting, i.e., using the measurements of a subset of qubits to adjust the global distribution. QuTracer \cite{QuTracer} further extends qubit subsetting by virtualizing Pauli Check Sandwiching \cite{gonzales2023quantum} to mitigate both gate and measurement errors. Furthermore, QEM is expected to complement error correction in the FTQC era \cite{FTQC1}\cite{FTQC2}\cite{FTQC3}\cite{zimborás2025mythsquantumcomputationfault}.

In addition to QEM methods, there are approaches that may not directly mitigate errors but instead transform them, thus making them easier to address. These techniques, known as noise-tailoring approaches, include: Dynamical Decoupling (DD) \cite{DynamicalDecoupling}, Twirled Readout Error eXtinction (TREX) \cite{PhysRevA.105.032620} and Pauli Twirling \cite{PauliTwirling}. DD reduces errors caused by decoherence when a qubit is idle. DD achieves this by adding strong rapid sequences of control pulses that do not affect the output but suppress decoherence-induced errors. For noise that is Markovian in nature, DD ``tailors" it to a symmetric noise channel, while for non-Markovian noise, DD can effectively decouple the system from the environment, leading to effective noise suppression \cite{DD-Mitiq}. Pauli Twirling converts errors such as decoherence and coherent errors into a randomized Pauli channel, making them easier to mitigate. 

In this study, 
we propose a novel QEM scheme named Q-Cluster to reshape noisy output probability distributions. We start with a simple bit-flip noise model, which implies that erroneous bit-strings tend to ``cluster'' around ideal ones based on their Hamming distance. Hence, Q-Cluster uses Hamming distance-based clustering to infer the structure of the output distribution. Qubit-wise Majority Vote (QMV) is used to determine the centroids of each cluster. QMV is chosen as it provides maximal likelihood estimation for a single correct output under i.i.d symmetric bit-flip noise. Additionally, QMV offers the benefit of recovering unobserved bit-strings from noisy results \cite{baron2024maximumlikelihoodquantumerror}. 
The clustering results are then used to reshape the output distribution in a noise-aware manner: (a) the noise impact is computed as the probability of a non-centroid bit-string occurring as a result of noise-induced bit-flips from centroids; and (b) the non-centroid bit-strings' probabilities are then lowered accordingly to reverse the noise impact. Our results show that this clustering approach is highly effective in mitigating bit-flip noise, especially when the number of clusters is known a priori. To handle the question with unknown numbers of clusters, we propose an iterative approach: the clustering process repeats with increasing numbers of clusters until there is diminishing return from re-shaping the output distribution.

For Q-Cluster to perform on real quantum hardware, where noise is more complex than simple bit-flips, we introduce two enhancements: (a) noise tailoring by applying Dynamic Decoupling (DD) and Pauli Twirling, and (b) machine learning (ExtraTrees) to estimate an effective bit-flip error rate. The overall effect is that the tailored noise can be approximated with a bit-flip noise model with the effective error rate. 

\begin{figure*}
    \centering
    \includegraphics[width=0.9\textwidth]{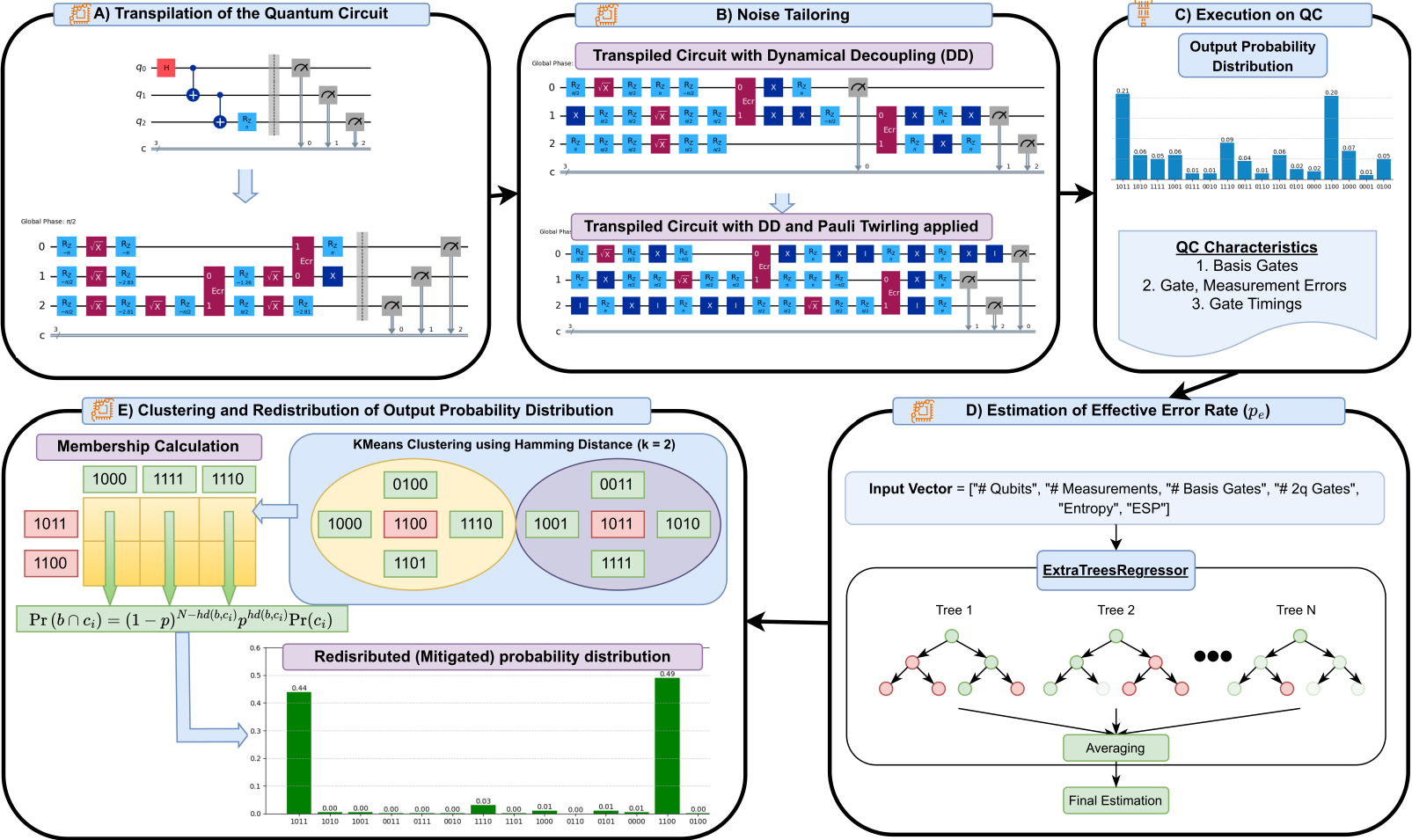}
    \caption{Q-Cluster workflow. Stage A: transpilation of the input circuit to the target quantum computer. Stage B:  application of noise tailoring techniques, Dynamical Decoupling and Pauli Twirling, to shape quantum noise into a local depolarizing model (Sec. \ref{sec:NoiseTailor}). Stage C: execution of the noise-tailored circuits on the target device, and the results, along with noise and timing characteristics, are collected. Stage D: Approximation of the tailored noise effects as a bit-flip noise model with an effective error rate  $p_e$ (Sec. \ref{subsec:pe}). Stage E: Noise mitigation in the output probability distribution using noise-aware clustering (Sec. \ref{sec:cluster}) and noise-aware redistribution (Sec. \ref{sec:redist}).}
    \label{fig:Q-Cluster_block}
\end{figure*}

Our experimental results on two recent generations of IBM Quantum processors demonstrate that our Q-Cluster approach improves the fidelity by 1.46x, and outperforms the state-of-the-art distribution reshaping algorithms. The workflow of our proposed Q-Cluster scheme is shown in Fig. \ref{fig:Q-Cluster_block}.

The remainder of the paper is organized as follows. Sec. \ref{sec:Background} provides background on related quantum error mitigation algorithms. 
 Sec. \ref{sec:Design} presents the design of Q-Cluster under the assumption of a bit-flip noise model. In Sec. \ref{sec:bit-flip}, we evaluate Q-Cluster on this noise model to demonstrate its effectiveness
and motivate further design decisions. Building on insights from Sec. \ref{sec:bit-flip}, Sec. \ref{sec:realdevice} incorporates noise tailoring and a machine learning–based effective error rate estimator to adapt Q-Cluster for real quantum hardware. Sec. \ref{sec:Experiments} evaluates Q-Cluster’s performance against state-of-the-art QEM methods on actual IBM quantum devices. Finally, Sec. \ref{sec:Conclusion} summarizes our findings and conclusions. 

\section{Background}\label{sec:Background}

QEM methods can be classified into two broad categories: those correcting expectation values (e.g., ZNE \cite{ZNE}, PEC \cite{PEC}) and those reshaping output distributions (e.g., M3 \cite{m3}, HAMMER \cite{Hammer}, and QBEEP \cite{qbeep}). Our proposed Q-Cluster belongs to the latter category.

Measurement noise affects a quantum state as $\vec{p}_{noisy} = A\vec{p}_{ideal}$, 
where $\vec{p}$ represents a $2^N$-dimensional probability vector, and $A$ is a $2^N \times 2^N$ matrix. Each element $a_{(i,j)}$ in the matrix indicates the probability that the noise changes the bit-string $j$ to the bit-string $i$. The matrix $A^{-1}$ can be applied to the noisy probability vector to correct it. M3 follows the matrix inversion approach to mitigate measurement errors, by constructing matrix $A$ through calibration runs on the quantum device. 

On the other hand, HAMMER assigns weights based on the probability distribution of the Hamming distance 
{$d \in [1, \lfloor N/2 \rfloor]$} around each bit-string in the output distribution. QBEEP builds a graph of the observed bit-strings, with bit-strings as nodes and weighted edges as the probability of re-classifying one bit-string to another. QBEEP uses the Poisson distribution and Bayesian inference to reassign probabilities between bit-strings. 

A common limitation among M3, HAMMER, and QBEEP is that their corrections can only be applied to the measured bit-strings. This limitation requires that the bit-strings of interest must have been measured or observed at least once. With high circuit depths or widths, the probability of measuring all the qubits correctly at the same time may be small, which in turn requires a large number of shots to satisfy this requirement.
Another important consideration for mitigation is that when noise is absent or minimal, the correction should also be negligible. M3 and QBEEP adjust their mitigation based on the noise error rates obtained from the machine. In contrast, HAMMER lacks this adaptability and applies mitigation even in the absence of noise, leading to potentially erroneous corrections.

\section{Q-Cluster Design} \label{sec:Design}


In this section, we detail the design of the Q-Cluster algorithm.
Sec.~\ref{sec:Design}.A presents the process of identifying dominant structures in the noisy distribution via clustering. Sec.~\ref{sec:Design}.B proposes a noise-aware probabilistic re-distribution to mitigate the noise effect.
Sec.~\ref{sec:Design}.C discusses our iterative approach to determine the number of clusters. Sec.~\ref{sec:Design}.D provides an illustrative example of Q-Cluster,

\subsection{Recovering Dominant Bit-Strings with Clustering}\label{sec:cluster}

A key observation or assumption behind Q-Cluster is that the dominant structure of a noisy distribution can be recovered with clusters. For simplicity, we start with a symmetric bit-flip noise model. With this model, the effect on a single qubit is:
\begin{equation}
|\psi_{\text{noisy}}\rangle = (1-p)|\psi\rangle + p X|\psi\rangle
\end{equation}

For $N$-qubits, the probability of $k$ bit-flips occurring is:

\begin{equation}
P(k) = \binom{N}{k} p^k (1-p)^{N-k}
\end{equation}

From this model, we can see the noise impact: the noisy distribution would be centered around the noiseless bit-string based on Hamming distance (HD),
as exemplified in Fig. \ref{fig:randbf-case-study}a (noiseless) and \ref{fig:randbf-case-study}b (noisy). Therefore, if the output of a quantum circuit has a few ``dominant'' bit-strings that are relatively separated in Hamming space, they can be well recovered using a clustering scheme: each cluster would represent a dominant bit-string in the noise-free distribution. Therefore, by identifying the salient feature, i.e., clusters, from the noisy distribution, we can recover the dominant patterns of the noise-free distribution.

We term the distributions with a limited number of dominant bit-strings as low entropy due to their low (Normalized) Shannon entropy, which is computed as: 
\begin{equation}\label{eqn:ShannonE}
    \text{Entropy (Normalized)} = \frac{- \sum_{i = 0}^n p(x_i)\text{log}_2 (p(x_i))}{\text{log}_2 2^N}
\end{equation} 

Q-Cluster uses the $K$-Means clustering algorithm \cite{macqueen1967some}. $K$-Means clustering requires the user to specify the number of clusters ($K$). For each cluster, $K$ centroids are initialized from the dataset, and the remaining data points are assigned to the nearest centroid. 
While Euclidean distance is typically used to quantify nearness or proximity, we adopt HD as our metric to quantify nearness or proximity.

 The next step in K-Means clustering is to identify the cluster centroids. Baron et al. \cite{baron2024maximumlikelihoodquantumerror} proved that under an i.i.d. (independent identical distribution) symmetric bit-flip noise model, QMV (quantum majority vote) computes the maximal likelihood estimate when there is a single correct bit-string in the noisy distribution. 
As we expect one centroid per cluster, QMV is the perfect choice. In other words, to compute the centroid of a cluster, a majority vote is taken for each qubit across all the shots belonging to the cluster. 


However, bit-strings that are outliers to a cluster can lead to an incorrect QMV result. To ensure that such outliers do not affect the majority vote, we need to exclude them from the clusters. We identify the outliers in the following manner. Given the bit-flip model, the variance of bit-flip errors is $\text{Var} = Np(1-p)$, where $N$ is the number of qubits and $p$ is the bit-flip error rate. This implies that bit-strings with a HD from the centroid greater than the standard deviation (i.e., the square root of the variance) are less likely to belong to the corresponding cluster. 
Hence, we impose a threshold while labeling the bit-strings into clusters. If the HD between the bit-string and the centroid it is assigned to, is too large (more than twice the variance $\theta = 2 \times Var$),
i.e., $HD > \theta$,
we label the bit-string as unassigned outliers. 
Excluding outliers prevents them from influencing the majority vote within a cluster. Additionally, it makes the clustering process noise-aware. In the noise-free case ($p = 0$), each bit-string becomes its own cluster centroid, leaving the distribution unchanged as $\theta = 0$. As $p$ increases, cluster sizes grow to reflect the increased spread of errors.

Note that by discarding outliers, the Q-Cluster behaves differently from QMV. Even in scenarios with a single correct output, Q-Cluster’s thresholding makes it more resilient to noisy outlier bit-strings, improving the robustness in high-noise regimes. 

\subsection{Re-Distribution of Bit-Strings}\label{sec:redist}

After discovering the clusters from a noisy distribution, we propose a method based on Bayesian inference to reshape the output probability distribution to undo the noise effect. 

Given the clustering results, centroids and their associated bit-strings, we assume that centroids are noise-free, while bit-strings away from the centroids are likely noise-induced. We also consider the possibility that a non-centroid bit-string can be a result of noise on different centroids based on its HD to these centroids. Therefore, we first compute the probability of non-centroid bit-strings arising from each centroid. Then, we reduce its overall probability to reverse the noise effect, as shown in Eqn.~\ref{eqn:rev},
\begin{equation} \label{eqn:rev}
\Pr(b)_{out} = \Pr(b)_{in} - \sum_{i} \Pr{(b \cap c_i)}.
\end{equation}
In the equation, $\Pr(b)_{in}$ is the probability of the bit-string $b$ from the noisy distribution, $\Pr(b)_{out}$ is the adjusted probability of the bit-string $b$, and $\Pr{(b \cap c_i)}$ is the joint probability of bit-string $b$ and centroid $c_i$, representing the probability of bit-string $b$ being observed while the ground truth is $c_i$.
This joint probability can be computed as follows. 
\begin{align}\label{eqn:membership}
\Pr(b \cap c_i) &= \Pr(b \mid c_i) \Pr(c_i) \notag \\
                &= (1 - p)^{N - \text{HD}(b, c_i)} \, p^{\text{HD}(b, c_i)} \Pr(c_i).
\end{align}
In the equation, the conditional probability, $\Pr(b|c_i)$, is the probability of the bit-string $b$ being observed given the condition that the ground truth is $c_i$. This conditional probability is computed using the bit-flip model, $(1-p)^{N-hd(b, c_i)} \cdot p^{hd(b, c_i)}$, where $N$ is the number of qubits and $hd(b, c_i)$ is the HD between the bit-string $b$ and the centroid $c_i$. $\Pr(c_i)$ is the ratio of the number of bit-strings belonging to cluster $i$ over the overall number of shots. 

The purpose of redistribution is to reverse the noise effect. If the bit-string $b$ existed in the noiseless distribution, and was not dominant, its probability in the noisy distribution would increase, depending on its Hamming distance from the cluster centroids. Conversely, if $b$ did not exist in the original distribution, its probability in the noisy distribution is completely due to bit-flips from the dominant bit-strings. Therefore, if $p_{out}(b) \leq 0$, the bit-string is removed from the output distribution. This is an improvement over M3, HAMMER, and QBEEP, which do not remove the erroneous bit-strings. 

\subsection{Iterative Discovery of K}
A key assumption in $K$-Means clustering is that the number of clusters ($K$) is known \textit{a priori}. However, this assumption is often impractical. 
To overcome this limitation, we propose an iterative approach to determine $K$. Starting with $K = 1$, we gradually increase the number of clusters. 
In each iteration, after we perform the output re-distribution, we check the Hellinger fidelity (HF) between the \( i^{\text{th}} \) iteration output \( p_{out} \) and the \( (i-1)^{\text{th}} \) iteration output \( p_{out} \) as HF quantifies the similarity between these two distributions. If the fidelity becomes larger than a predefined threshold $\delta$, we stop the iterative process and output \( (i-1)^{\text{th}} \) \( p_{out} \). Otherwise, we increase the number of clusters and repeat the procedure until: (1) we either satisfy this converging condition or (2) the number of clusters exceeds the number of unique bit-strings in the noisy distribution. 

The Q-Cluster algorithm is shown in Alg. \ref{alg:qcluster}.
\begin{algorithm}
\caption{Q-Cluster Algorithm} \label{alg:qcluster}
\KwIn{Noisy distribution $R_{noisy}$, error rate $p$, convergence threshold $\delta$, $K_{max} = \text{\# of unique bit-strings in} R_{noisy}$}
\KwOut{Mitigated distribution $R_{mitig}$}

Initialize number of clusters $K \leftarrow 1$ \;
Initialize mitigated distribution $R_{{mitig}}$ \;

\While{not converged and $K < K_{max}$}{
    
    \tcp{Step 1: Clustering}
    Select top-$K$ bit-strings as initial centroids \;
    Repeat until centroids stabilize:
    \begin{itemize}
        \item Assign bit-strings to nearest centroid based on HD
        \item Filter outliers using threshold $HD > \theta$
        \item Update centroids using majority vote
    \end{itemize}
    
    \tcp{Step 2: Redistribution}
    Adjust probabilities of bit-strings based on their proximity to centroids and the error rate ($p$)\;
    Re-normalize the output distribution \;
    
    \tcp{Step 3: Convergence Check}
    If  $\text{HF}(R_{mitig_K}, R_{mitig_{K-1}} ) > \delta$, break \;
    Otherwise, increment $K$ and repeat \;
}

Return final mitigated distribution ($R_{mitig_{K-1}}$)\;
\end{algorithm}
\subsection{An Illustrative Example}

\begin{figure}[h]
\centering 
\includegraphics[width=0.95\linewidth]{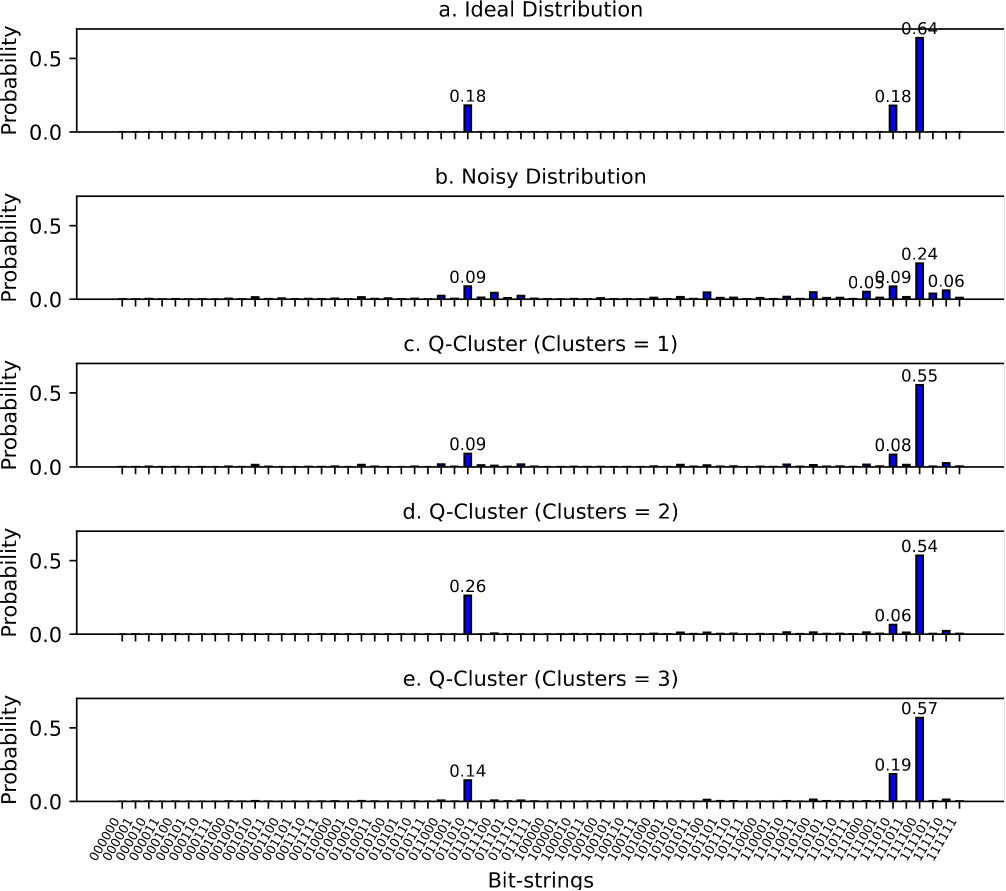} 
\caption{Illustrative example of Q-Cluster. (a) shows the ideal distribution, while (b) shows the noisy distribution after applying a bit-flip noise with error rate of p = 0.15. The noisy distribution serves as input to Q-Cluster. (c) shows the result after the first iteration of Q-Cluster, where the number of clusters (k) is 1. (d) and (e) show the resulting distributions of Q-Cluster for k = 2 and k = 3, respectively. (Bit-strings are sorted according to Hamming distance from `000000')} 
\label{fig:randbf-case-study} 
\end{figure}

To illustrate how our Q-Cluster algorithm works, we present a case study in Fig. \ref{fig:randbf-case-study}. The noise-free probability distribution ($R_{ideal}$) contains three randomly generated dominant bit-strings (`111000', `111010', `011010') with different probabilities as shown in Fig. \ref{fig:randbf-case-study}a. We apply a random bit-flip with the probability of $p = 0.15$ to $R_{ideal}$, resulting in a noisy distribution $R_{noisy}$ shown in Fig. \ref{fig:randbf-case-study}b. The Q-Cluster algorithm uses $R_{noisy}$ as input, along with the bit-flip error rate ($p = 0.15$) information. We set the iterative clustering stopping threshold ($\delta$) to 0.9.

Fig. \ref{fig:randbf-case-study} c through e shows the result of each iteration during the Q-Cluster process. Q-Cluster begins with $K = 1$ cluster. It selects the bit-string `111000' as the initial centroid (Line 4 in Alg. \ref{alg:qcluster}), as it has the highest probability in the noisy distribution. The algorithm then performs clustering to identify the cluster population, followed by QMV (Line 5 in Alg. \ref{alg:qcluster}). 
Since `111000' is the only cluster and the variance of the cluster is $Var = Np(1-p) = 0.765$, the outlier identification threshold ($2\times Var$) is $\theta =  \lceil 1.53 \rceil = 2$. 
Therefore, only bit-strings, whose HD is less than 2 from the centroid, are assigned to the cluster. After the bitwise majority vote of cluster members, the centroid remains `111000'. Then, during the redistribution step (Line 6-7 in Alg. \ref{alg:qcluster}), the probability of the bit-strings close to `111000' is reduced. Meanwhile, the bit-string `011010', which is $HD = 2$
away from the centroid, largely retains its probability from the noisy distribution. 

In the next iteration, with $K = 2$ clusters, Q-Cluster adds `011010' as a new centroid. After re-distribution, the probability of `011010' nearly triples, as shown in Fig. \ref{fig:randbf-case-study}c, because many bit-strings that were previously distant from the `111000' cluster now fall within the `011010' cluster. The bit-string `111010', which is equidistant from both centroids, has its probability reduced by 33\% as some of it is now reassigned to the second cluster.

As the fidelity between the $K = 1$ and $K = 2$ distributions is 0.85 — below the stopping threshold of 0.90 — the algorithm proceeds to $K = 3$ (Line 8-9 in Alg. \ref{alg:qcluster}). During this iteration, Q-Cluster adds `111010' as a new centroid. Since it lies between the two previous centroids, it absorbs parts of their surrounding bit-strings during redistribution. The fidelity between the $K = 2$ and $K = 3$ distributions increases to 0.89, prompting one more iteration.

The clustering continues up to $K = 4$ (not pictured), but the addition of a new cluster center does not significantly alter the distribution (fidelity exceeds 0.90). This indicates that adding a new cluster was unnecessary. Consequently, Q-Cluster outputs the distribution observed at $K = 3$, Fig. \ref{fig:randbf-case-study}e, as the final result (Line 10 in Alg. \ref{alg:qcluster}). As we can see from the figure, the noise-mitigated distribution from Q-Cluster, Fig. \ref{fig:randbf-case-study}e, is very close to the noise free distribution, \ref{fig:randbf-case-study}a, compared to the noisy distribution \ref{fig:randbf-case-study}b. 


\section{Q-Cluster on Bit-Flip noise model} \label{sec:bit-flip}

In this section, we first show that Q-Cluster remains effective in mitigating bit-flip noise even at high error rates.
We then conduct a parameter sensitivity analysis to evaluate how sensitive Q-Cluster is w.r.t. the stopping threshold ($\delta$) and the bit-flip error rate ($p$) estimation. 
Finally, we explore how the prior knowledge of the number of dominant bit-strings in the ideal distribution influences the effectiveness of Q-Cluster.

\subsection{Mitigation Effectiveness on Bit-flip Noise Model}
To quantify mitigation effectiveness, we use improvement in fidelity\cite{qbeep}, which is computed as follows,
\begin{equation}\label{eqn:Improvement}
Improvement = \frac{HF_{mitigated} + \epsilon}{HF_{noisy}+ \epsilon} ,
\end{equation}
where HF is the Hellinger fidelity between the noisy ($HF_{noisy}$) or mitigated ($HF_{mitigated}$) probability distribution and the noise-free probability distribution, and $\epsilon$ is a regularization constant (we use 0.01) for computing the geometric means across benchmarks.

An improvement value less than or equal to 1 indicates that the mitigation did not enhance fidelity or made it worse, while a value greater than 1 indicates beneficial mitigation. 

\begin{figure*}
    \centering 
\includegraphics[width=0.95\textwidth]{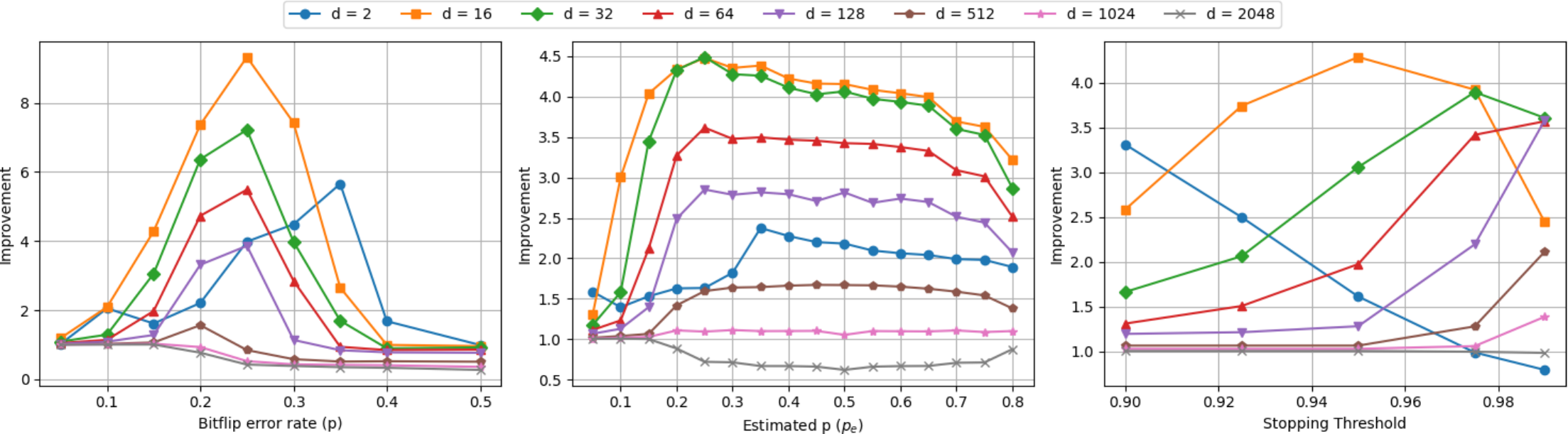} 
\caption{Q-Cluster fidelity improvement with different parameters (Higher Improvement; better mitigation). Each point on the plot shows the average result over 10 randomly generated probability distributions, where the number of qubits is 14 and $d$ is the number of dominant states in the ideal distribution. (a) Effect of  different bit-flip error rates.  (b) Effect of mis-estimated bit-flip error rate $p$ (actual p = 0.2). (c) Effect of different stopping thresholds..} 
\label{fig:Sensitivity Analysis} 
\end{figure*}
To understand where and why Q-Cluster is effective, we perform an analysis with bit-flip noise. In the experiments, we fixed the number of qubits to 14 and changed the number of dominant states ($d$), the bit-flip error rate ($p$), and the threshold ($\delta$). For each combination, we run 10 different randomly generated probability distributions, and report the mean improvement for each such unique combination in Fig. \ref{fig:Sensitivity Analysis}. 

\subsubsection{Effectiveness of Q-Cluster for Different Bit-flip Error Rates}

Fig. \ref{fig:Sensitivity Analysis}a shows that Q-Cluster performs well for nearly all $d$ values with up to a bit-flip error rate of $p = 0.15$, beyond which only distributions with $d < 512$ ($Entropy < 0.64$) exhibit improvements for $p$ up to $0.3$. Beyond this, only marginal improvements are observed for $p$ up to $0.4$, and only for low-$d$ cases. The figure also highlights that Q-Cluster is particularly effective for low-entropy systems ($Entropy < 0.5$ or $d < 128$), even at high error rates such as $p = 0.4$.

To contextualize the severity of $p = 0.4$: for $N = 14$ and $d = 1$, the probability of measuring a correct bit-string is approximately $(1 - 0.4)^{14} = 0.0007$. Even under such extreme noise, Q-Cluster achieves fidelity improvements greater than 1.5.
This robustness is due to the clustering mechanism in Q-Cluster. While the state-of-the-art methods like QBEEP, HAMMER, and M3 also perform redistribution, Q-Cluster first identifies clusters of bit-strings that have likely resulted from bit-flips of the original dominant strings. It then uses the amplified probability mass of these clusters to guide redistribution more accurately, thereby achieving stronger error mitigation in high-noise regimes.

\subsubsection{Sensitivity of Q-Cluster to Inaccurate Effective Bit-flip Error Rates}
Up to this point, we have assumed prior knowledge of the bit-flip rate. However, in 
practice, this is rarely the case, and we must estimate this error rate, referred to as {\em effective error rate}, $p_e$. To explore what happens if we estimate $p_e$ inaccurately, we simulate a 14-qubit system with varying numbers of dominant states (d), subjected to bit-flip noise with a true probability of $p = 0.20$ per qubit. We then provide Q-Cluster with deliberately mis-estimated bit-flip probabilities ranging from $p_e = 0.05$ to $p_e = 0.35$, in steps of 0.05. The fidelity improvements of Q-Cluster are shown in Fig. \ref{fig:Sensitivity Analysis}b.

One might expect the improvement would peak when $p_e = p$ and decline as $p_e$ deviates from the true $p$ in either direction. However, as shown in Fig. \ref{fig:Sensitivity Analysis}b, the improvement actually peaks around $p_e = 0.25$ and remains relatively stable with only a slight decrease for $p_e > p$. In contrast, underestimating $p_e$ leads to a noticeable drop in performance. These results
suggest that overestimating $p$ is more favorable than underestimating it. Overall, Q-Cluster remains robust to estimation inaccuracies as long as $p_e > p$.

The reason for such fidelity improvement trend of Q-Cluster can be attributed to two main factors: (1) When $p_e < p$, the estimated variance, $Var = Np_e(1 - p_e)$, becomes too small, leading to overly small clusters due to Q-Cluster’s outlier identification ($HD > 2 \cdot Var$); and (2) a low $p_e$ results in insufficient re-distribution
of bit-strings, leaving many incorrect strings—those not present in the ideal distribution—untouched. On the other hand, when $p_e \geq p$, the clusters are appropriately sized, and even if $p_e$ is overestimated, Q-Cluster effectively eliminates incorrect bit-strings while preserving the correct proportions of dominant states.

\subsubsection{Sensitivity of Q-Cluster to the Stopping Threshold $\delta$}

The stopping threshold $\delta$ determines when Q-Cluster terminates its iterative process, i.e., when $HellingerFidelity(R_i, R_{i-1}) < \delta$, where $R_i$ is the mitigated distribution at iteration $i$. A higher $\delta$ generally results in more iterations (and more clusters) before convergence. While it may seem that increasing $\delta$ improves mitigation, Fig. \ref{fig:Sensitivity Analysis}c shows this is not universally true.

For low-entropy inputs (e.g., $d = 2$ and $d = 16$), higher $\delta$ values degrade 
performance, as Q-Cluster is forced to over-cluster in search of high fidelity between successive iterations. Conversely, for distributions with many dominant states, performance improves with higher $\delta$, as additional clusters capture finer structures.

These results suggest tuning $\delta$ based on expected cluster count: for low-cluster distributions ($d < 16$), $\delta \in [0.90, 0.93]$ is optimal; for high-cluster cases, $\delta \in [0.97, 0.99]$ yields better performance. When prior knowledge is unavailable, setting $\delta = 0.95$ provides robust performance across a wide range of scenarios.

\subsection{Q-Cluster with A Priori Numbers of Clusters}

Q-Cluster is designed to work without needing the user to specify the number of clusters in advance. However, if the user has prior knowledge of the number of clusters or an estimate of the number of dominant bit-strings, how would Q-Cluster perform in that case?

To answer the question, we perform the following study:
	1.	We provide the exact number of clusters to Q-Cluster and disable its iterative procedure for determining the cluster count.
	2.	We test the scenarios where the supplied number of clusters is under- or overestimated by 25\% and 50\% from the ground truths, respectively.
	3.	The performance in each case is compared against the default iterative version of Q-Cluster, which automatically determines the number of clusters based on a stopping threshold.


\begin{figure}
    \centering
    \includegraphics[width=0.9\columnwidth]{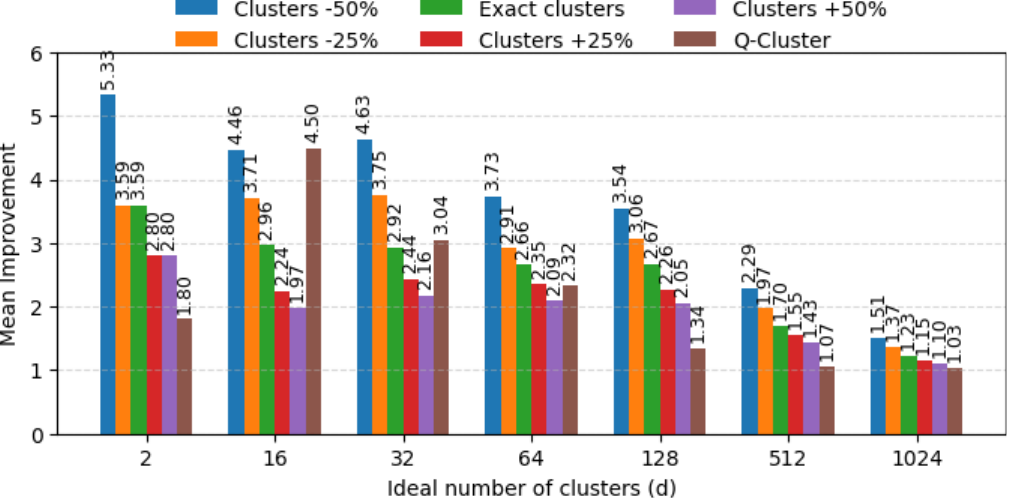}
    \caption{Performance of Q-Cluster when the user provides the expected number of clusters (Exact Clusters). The experimental set up is the same as used in Fig. \ref{fig:Sensitivity Analysis} with $p = 0.15$. We also test the effects under-estimating the number of clusters by 25\% and 50\% (Clusters -25\% and Clusters -50\%) or overestimating the number of clusters by same amount (Clusters +25\% and Clusters +50\%). We contrast this with Q-Cluster's iterative Cluster finding approach.}
    \label{fig:clus_sens}
\end{figure}
Fig. \ref{fig:clus_sens} presents the results of our investigation. It shows that, except for some low-entropy cases ($d = 16$ or $d = 32$), Q-Cluster would benefit from user-provided numbers of clusters even when they are underestimated or overestimated for certain degrees. However, underestimating the number of clusters consistently outperforms overestimation.

Overestimating often introduces spurious centroids, pulling probability mass away from correct clusters. In contrast, underestimation still preserves dominant bit-strings as centroids, boosting their probabilities and eliminating some erroneous strings. This is also illustrated in Fig. \ref{fig:randbf-case-study} 
 when Clusters = 1, i.e., the number of clusters is underestimated by 66\%, yet correct bit-strings remain intact while many incorrect ones are suppressed.

\section{Q-Cluster on Real Quantum Devices} \label{sec:realdevice}

Real devices exhibit complex noise that deviates from the bit-flip model. Instead of modeling all noise sources, we use noise tailoring to make the noise behave more similarly to bit-flips. Then, to best capture the tailored noise effect, we leverage machine learning (ML) to estimate the effective error rate $p_e$.

\subsection{Noise Tailoring}\label{sec:NoiseTailor}
A real hardware device has multiple noise sources such as coherent, crosstalk, and decoherence errors. Many have a non-linear noise channel. For example, the decoherence error is given by:
$P(t) = (1 - e^{-t/T_1})P(0)$. 
Such noise does not fit the random bit-flip model because of its dependence on time. 

To make the noise closer to a bit-flip noise model , we resort to noise-tailoring techniques. Dynamical decoupling (DD) \cite{DynamicalDecoupling} helps reduce the effects of decoherence noise. Pauli twirling \cite{PauliTwirling} converts complex error sources into a randomized Pauli channel. 
DD and Pauli twirling are complementary to each other and to Q-Cluster. These techniques ``tailor" the noise in a way that makes it easier for Q-Cluster to mitigate, without requiring additional circuit runs.

\subsection{Estimating Effective Error Rate} \label{subsec:pe}


Fig. \ref{fig:Sensitivity Analysis} demonstrates Q-Cluster's effectiveness on bit-flip noisy distributions, assuming prior knowledge of the bit-flip error rate $p$. In practice, this assumption is unrealistic, as $p$ varies across circuits, devices, and due to the stochastic nature of noise. Therefore, estimating the effective error rate for a given circuit is essential.

Common error estimation techniques include Estimated Success Probability (ESP) \cite{ESP}, Probability of Successful Trials (PST), QuEST \cite{QuEST}, and Clifford-based fidelity estimation \cite{seifert2024claptoncliffordassistedproblemtransformation}. These methods have been incorporated into various mitigation frameworks, such as RZNE \cite{RZNE}, QRAFT \cite{QRAFT}, and QEM using noise estimation circuits \cite{urbanek2021mitigating}.
PST and Clifford-based methods require additional quantum circuits, whereas ESP offers a non-intrusive alternative. ESP is computed from calibration data, using gate and measurement error rates.


The issue with these existing error rate estimation approaches is that the estimated error rate is not based on a bit-flip noise model. This issue arises from two main reasons. First, the bit-flip model assumes independent and identically distributed (i.i.d.) errors across qubits, whereas in practice, qubits with higher gate depths typically experience greater error accumulation. Second, these estimation approaches often fail to capture complex noise sources such as crosstalk, idling errors, and over/under rotation, leading to inaccuracies. To address the issue, we propose to use a machine learning model to estimate the effective bit-flip error rate, that takes all the aforementioned errors into account.  

\begin{table}[h]
    \centering
        \caption{Mean Squared Error (lower the better) and $R^2$ (higher the better) comparison of different ML models for estimating $p_e$. The MSE and $R^2$ values are the averages across five fold cross validation.}
    \begin{tabular}{|l|c|c|}
        \hline
        Model & Average MSE & Average $R^2$ \\
        \hline
        RandomForest \cite{RandomF} & 0.0006 & 0.9535 \\
        GradientBoosting \cite{GradBoosting} & 0.0006 & 0.9517 \\
        \textbf{ExtraTrees} \cite{ExtraTrees} & \textbf{0.0005} & \textbf{0.9643} \\
        DecisionTree \cite{DecTree} & 0.0010 & 0.9229 \\
        SVR \cite{drucker1997support} & 0.0126 & -0.0199 \\
        KNN \cite{KNN} & 0.0026 & 0.8014 \\
        XGBoost \cite{chen2016xgboost} & 0.0006 & 0.9452 \\
        \hline
    \end{tabular}
    
    \label{tab:model_performance}
\end{table}

To identify the most effective machine learning model for estimating the effective error rate $p_e$, we evaluated seven regression models using five-fold cross-validation, comparing their mean squared error (MSE) and $R^2$ scores. The dataset comprised 171 circuits executed on five quantum machines: \textit{ibm\_strasbourg}, \textit{ibm\_brisbane}, \textit{ibm\_torino}, \textit{ibm\_brussels}, and \textit{ibm\_kyiv}. As reported in Table \ref{tab:model_performance}, the ExtraTrees Regressor achieved the best performance, yielding the lowest MSE and highest $R^2$ score.
Table \ref{tab:feature_sources} lists the features used for $p_e$ estimation along with their sources. All features were extracted from either the transpiled circuit or the backend calibration data, requiring no additional quantum circuit executions.
To generate training labels for $p_e$, we leveraged the availability of both ideal and noisy output distributions for each circuit and assumed an i.i.d. bit-flip noise model. Under this model, the noisy probability of a bit-string $b$ is given by:

\begin{equation} \label{eqn:prbnoisy}
Pr(b)_{\text{noisy}} = Pr(b)_{\text{ideal}} \cdot (1 - p_e)^N 
\end{equation}
where $N$ is the number of measured qubits. Rearranging Eq. \ref{eqn:prbnoisy} yields an expression for the effective bit-flip error rate:

\begin{equation} \label{eqn:prbest} p_e = 1 - \sqrt[N]{\frac{Pr(b)_{\text{noisy}}}{Pr(b)_{\text{ideal}}}} \end{equation}

\begin{table}[htb]
    \centering
    \renewcommand{\arraystretch}{1.1} 
    
    \caption{Features used to estimate $p_e$, their descriptions and sources.}
    \begin{tabular}{|p{0.21\columnwidth}|p{0.41\columnwidth}|p{0.23\columnwidth}|}
        \hline
        \textbf{Feature} & \textbf{Description} & \textbf{Source} \\ 
        \hline
        \# Qubits & Number of active qubits & Transpiled circuit \\ 
        \# Measurements & Number of measured qubits & Transpiled circuit \\ 
        \# 2q gates & Number of two-qubit gates & Transpiled circuit \\ 
        \# SX gates & Number of SX & Transpiled circuit \\ 
        \# X gates & Number of X  & Transpiled circuit \\ 
        \# RZ gates & Number of RZ & Transpiled circuit \\ 
        Entropy & Shannon entropy of $R_{noisy}$ & Quantum Device \\ 
        ESP & ESP as calculate in ref \cite{ESP}& Quantum Device \\ 
        \hline
    \end{tabular}
    \label{tab:feature_sources}
\end{table}

\begin{figure}
    \centering
    \includegraphics[width=0.7\columnwidth]{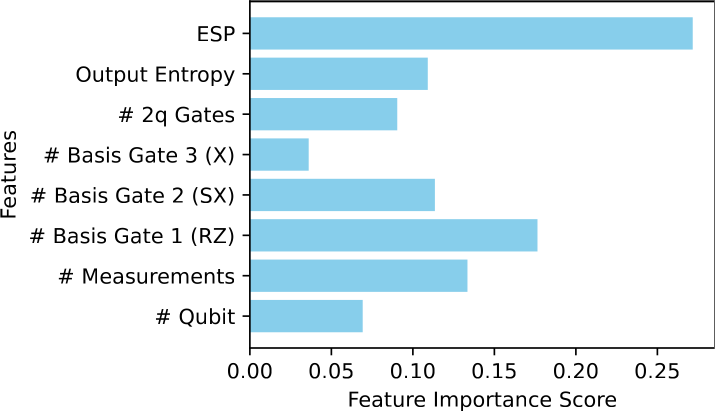}
    \caption{The feature importance scores of the trained ExtraTreesRegressor model, showing the relative significance of each feature in estimating the effective error rate  $p_e$. Among them, ESP has the highest importance, followed by the number of measurements, the number of qubits, and entropy. }
    \label{fig:FeatureImportances}
\end{figure}

Fig. \ref{fig:FeatureImportances} presents the feature importance learned by the model. The most important feature is ESP, which aggregates single-qubit gate, two-qubit gate, and measurement error rates, thereby accounting for the primary sources of noise in quantum circuits. This is followed by the number of measured qubits and total qubits. These two numbers are treated separately as they can differ in certain circuits (e.g., adders or Bernstein-Vazirani), where not all qubits are measured. Another significant feature is the entropy of the output distribution. High-entropy outputs are inherently more resilient to bit-flip noise than low-entropy ones when measured in the computational basis. Thus, if both ESP and output entropy are high, then the noise-free distribution would likely be high-entropy. Conversely, if ESP is low but the output entropy is high, this indicates a low-entropy noise-free distribution that has been heavily degraded by noise, thereby implying a high $p_e$. The machine learning model captures these patterns effectively, thereby improving Q-Cluster’s performance.
 \begin{figure}[htb]
     \centering
     \includegraphics[width=0.9\columnwidth]{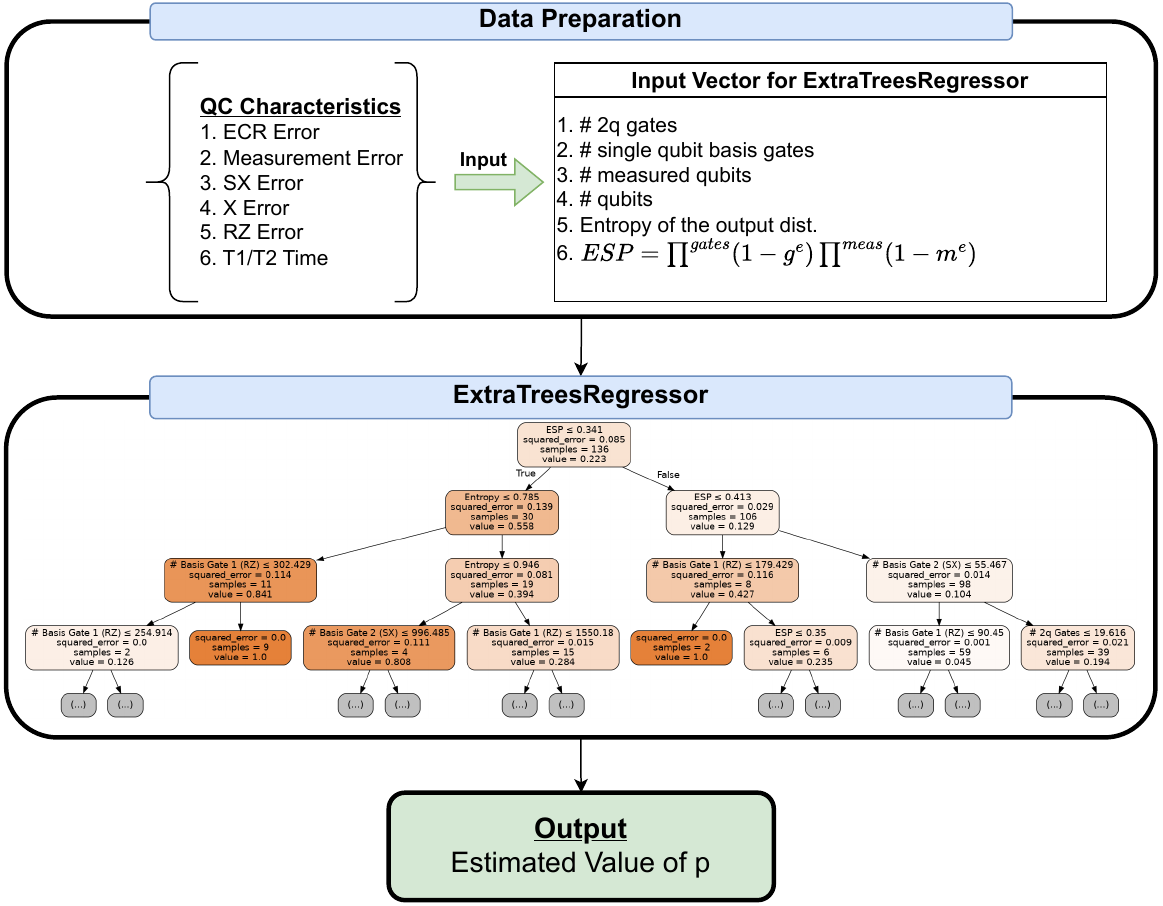}
     \caption{The procedure for estimating the effective error rate  $p_e$  using ExtraTreesRegressor.
     }
     \label{fig:ExtraTrees}
 \end{figure}

A natural question is whether the trained machine learning model generalizes to unseen devices and how frequently it needs retraining. To evaluate this, we trained an ExtraTreesRegressor on circuits executed on IBM Eagle r1 architecture devices: \textit{ibm\_strasbourg}, \textit{ibm\_brisbane}, \textit{ibm\_brussels}, and \textit{ibm\_kyiv}. We then tested the model on \textit{ibm\_torino}, which features the Heron r1 architecture. Notably, while both architectures share the same single-qubit basis gates, they differ in two-qubit gate implementations: Eagle uses ECR gates, whereas Heron employs the CZ gates. Heron reportedly has very low crosstalk errors.
The results of this evaluation are summarized in Table \ref{tab:torino_dataset}.
From the results in Table \ref{tab:torino_dataset} we can see that our regression model performs well even when not trained for the specific system. 

The procedure of using the ExtraTreesRegressor to infer the effective error rate, $p_e$, is shown in Fig. \ref{fig:ExtraTrees}

\begin{table}[h]
    \centering
    \caption{Comparison of Mean Squared Error (MSE), R2 Score, and Improvement with and without \textit{ibm\_torino} in the training dataset.}
    \begin{tabular}{|l|>{\centering\arraybackslash}m{2cm}| >{\centering\arraybackslash}m{2.5cm}|}
        \hline
        & \makecell{Trained with\\ \textit{ibm\_torino}} & \makecell{Trained without \\ \textit{ibm\_torino}} \\
        \hline
        MSE          & 0.0005  & 0.0009 \\
        R2 Score     & 0.9643  & 0.8858 \\
        \hline
    \end{tabular}
    
    \label{tab:torino_dataset}
\end{table}

\subsection{Time Complexity of Q-Cluster}
\begin{table}[h]
\centering
\caption{Worst-Case Time complexity comparison}
\begin{tabular}{|c|c|c|c|}
\hline
\textbf{M3} & \textbf{HAMMER} & \textbf{QBEEP} & \textbf{Q-Cluster} \\ \hline
$O(Ns)$ & $O(s^2)$ & $O(Ns^2)$ & $O(s^2)$\\ \hline
\end{tabular}

\label{tab:1}
\end{table}
The most computationally intensive part of Q-Cluster is the reassignment of bit-strings. This process scales with the number of bit-strings, $b$, multiplied by the number of centroids, $c$, resulting in a complexity of $O(b \cdot c)$. In the worst-case scenario, the number of bit-strings can equal the number of shots taken, and the number of centroids can also equal the number of shots ($s$). Thus, the worst-case complexity for our algorithm is $O(s^2)$.
In comparison, QBEEP reports its worst-case complexity as $O(Nr)$, where $r$ is the number of edges in their correction graph and $N$ is the number of qubits. In QBEEP's worst case, the number of nodes can equal the number of shots $s$, and the number of edges would be $r = s^2$. Table \ref{tab:1} summarizes the worst-case time complexity of each method.
These complexities represent the worst case. Practical scenarios often result in much lower values, as reported in Sec. \ref{sec:runtime_res}. An important note is that all the computations can be parallelized for much quicker execution.


\section{Experiments and Results}\label{sec:Experiments}

\subsection{Methodology}
In our experiments, we test our Q-Cluster as well as the state-of-art QEM methods, M3 \cite{m3}, HAMMER \cite{Hammer}, and QBEEP \cite{qbeep}. We also report the performance of Q-Cluster variants where we overestimate $p_e$ by 10\% and 50\%, denoted by ``Q-Cluster + 10\%'' and ``Q-Cluster + 50\%'', respectively. We test these methods on 29 benchmarks chosen from QASMBench \cite{QASMBench} and SupermarQ \cite{Supermarq}. Specifically, the benchmarks are chosen from the QASMBench-small, QASMBench-medium, and Supermarq benchmarks of Hamiltonian Simulation, VQE, and Vanilla QAOA. Table \ref{tab:benchmarks_entropy} lists the benchmarks used in our study. The low entropy benchmarks are the benchmarks whose noise-free Shannon entropy is less than 0.6, whereas the high entropy benchmarks have a Shannon entropy higher than 0.6.

\begin{table}[h]
\centering
\caption{Benchmarks}
\begin{tabular}{|c|p{6.2cm}|}
\hline
\textbf{Type} & \textbf{Benchmarks} \\ \hline
\textbf{Low Entropy} & adder\_n10, adder\_n4, basis\_change\_n3, basis\_test\_n4, basis\_trotter\_n4, bv\_n14, bv\_n19, cat\_state\_n22, cat\_state\_n4, fredkin\_n3, ghz\_state\_n11, ghz\_state\_n23, hamsim\_20, hs4\_n4, linearsolver\_n3, lpn\_n5, qec\_en\_n5, sat\_n7, toffoli\_n3, variational\_n4, wstate\_n3, wstate\_n27 \\ \hline
\textbf{High Entropy} & hamsim\_10, qaoa\_10, qaoa\_15, qft\_n4, qrng\_n4, vqe\_10, vqe\_20 \\ \hline
\end{tabular}
\label{tab:benchmarks_entropy}
\end{table}
For the Extra Trees Regressor, we used the implementation provided in the scikit-learn \cite{sklearn_api}. The dataset consists of a total of 145 circuits. We used a five-fold Cross-Validation (CV) to obtain the MSE and $R^2$ scores. To ensure fairness, while estimating $p_e$, we ensure that the circuits for which $p_e$ is being estimated are not in the training set. 

We ran the benchmarks with DD and Pauli Twirling enabled on the IBM Quantum computers \textit{ibm\_brisbane}, \textit{ibm\_torino}, \textit{ibm\_brussels}, \textit{ibm\_kyiv}, and \textit{ibm\_strasbourg} (i.e., DD and Pauli Twirling have been incorporated in the baseline unmitigated results as well as all QEM results). To demonstrate the effectiveness of noise tailoring, we also executed the QEM approaches with and without Pauli Twirling. 
The performance metric is the fidelity improvement as per Eq. \ref{eqn:Improvement}.


Among the QEM methods, we used the mthree 
Python package for M3, HAMMER was implemented from the algorithm in \cite{Hammer}, and QBEEP is the vanilla implementation from the github repository provided by the authors \cite{QBeep-code}, we do not perform any parameter changes or optimizations. Our Q-Cluster code is open-sourced on github \cite{Patil_Q-Cluster_Quantum_Error}.

As discussed in Sec. \ref{sec:bit-flip}, clustering is not effective for mitigating distributions with high entropy. For these high-entropy benchmarks, we can either resort to M3 for mitigation or simply skip QEM for mitigating distribution as none of the QEM approaches provide a noticeable improvement.

\subsection{Effect of Pauli Twirling on Q-Cluster}

Noise tailoring is essential to Q-Cluster, which is specifically designed to mitigate bit-flip noise. 
Empirically, Q-Cluster with and without Pauli Twirling (PT) show a 41\% difference in fidelity improvement, demonstrating the importance of noise tailoring. 

To isolate which component of Q-Cluster benefits most from PT, we repeated the same experiments on \textit{ibm\_torino} and \textit{ibm\_strasbourg}. 
Specifically, we evaluated whether PT affects the accuracy of $p_e$ estimation. The circuits, calibration data, and backends were ensured to be identical, by running the PT and non-PT jobs consecutively. 
Note that these jobs were run separately from those in Sec. \ref{subsec:res-mach}. Any differences in $p_e$ estimation accuracy (5-fold cross-validation avg MSE) may reflect this.

\begin{table}[htb]
\centering
\caption{Comparison of Q-Cluster modules with and without Pauli twirling.}
\label{tab:twirling_comparison}
\resizebox{\columnwidth}{!}{%
\begin{tabular}{|c|c|c|c|c|}
\hline
\textbf{Q-Cluster Module} & \textbf{Metric} & \textbf{\begin{tabular}[c]{@{}c@{}}Twirling \\ Disabled\end{tabular}} & \textbf{\begin{tabular}[c]{@{}c@{}}Twirling \\ Enabled\end{tabular}} & \textbf{Notes} \\ \hline
\multirow{2}{*}{\textbf{ExtraTrees Regressor}} & MSE & 0.0518 & 0.0018 & Lower the better \\ 
 & R2 & 0.2692 & 0.8227 & Higher the better \\ \hline
\textbf{\begin{tabular}[c]{@{}c@{}}Clustering and \\ Re-Classification\end{tabular}} & Fidelity Improvement & 0.8134 & 1.1993 & Higher the better \\ \hline
\end{tabular}%
}
\end{table}

As shown in Table \ref{tab:twirling_comparison}, $p_e$ estimations on the same data are approximately 28x 
more accurate when PT is applied compared to when it is not. This indicates that the noise tailoring effect of PT directly enhances the accuracy of $p_e$ estimation, and not merely the downstream clustering performance.

We further evaluated whether Q-Cluster’s clustering and redistribution stages benefit from Pauli Twirling (PT) by supplying the exact value of $p_e$ (computed from Eq. \ref{eqn:prbest}). If PT impacts only the accuracy of $p_e$ estimation, then supplying the true $p_e$ should yield similar performance with or without PT. However, as shown in Table \ref{tab:twirling_comparison}, enabling PT results in a 1.47× improvement even with the exact $p_e$, indicating that PT simplifies mitigation by shaping the output distribution to more closely resemble a bit-flip noise profile.

\begin{figure}[htb]
    \centering
    \includegraphics[width=0.9\linewidth]{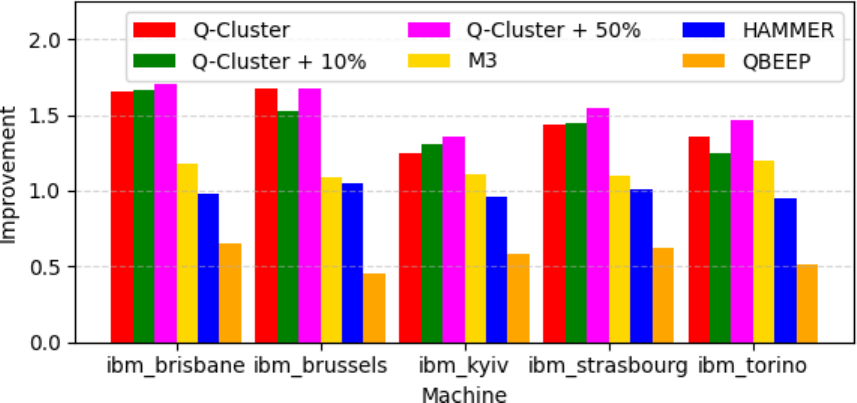}
    \caption{Comparison of QEM methods' mean improvement in fidelity on different IBM Quantum devices for low-entropy benchmarks.}
    \label{fig:overall}
\end{figure}

\subsection{QEM Methods on Different Quantum Computers} \label{subsec:res-mach}

\begin{figure*}[h!]
    \centering
    \includegraphics[width=0.9\linewidth]{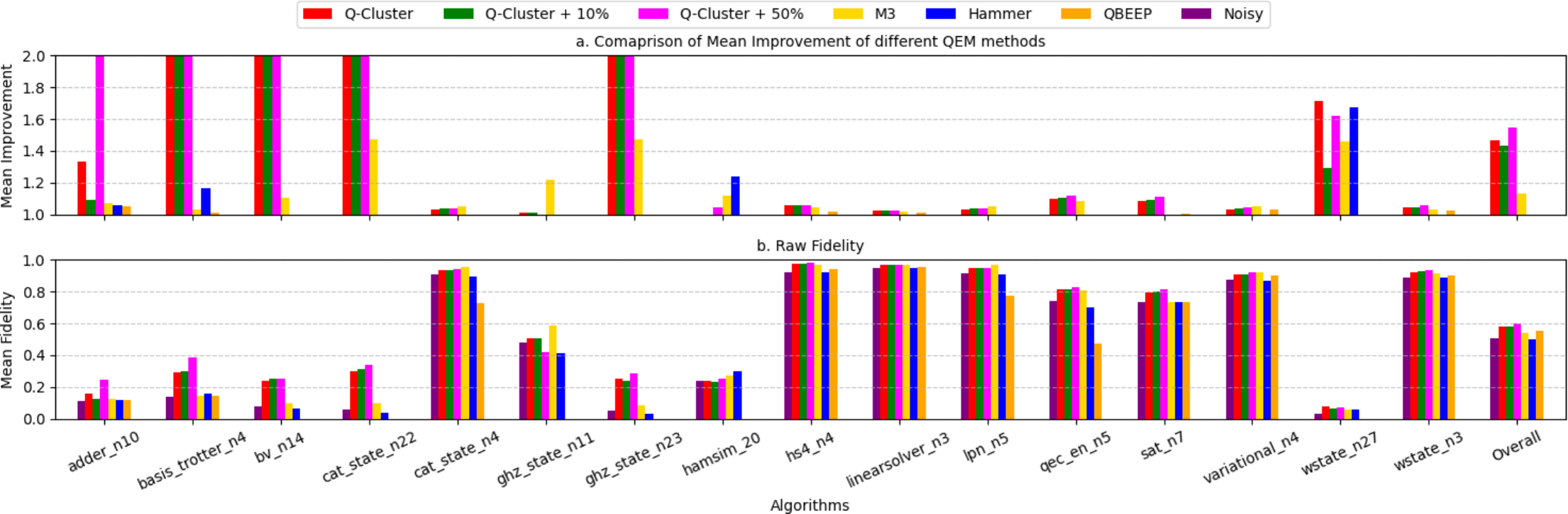}
    \caption{Comparison of raw fidelity and improvement in fidelity for QEM methods across IBM Quantum devices on low-entropy benchmarks.}
    \label{fig:raw}
\end{figure*}
\begin{figure}
    \centering
    \includegraphics[width=0.9\columnwidth]{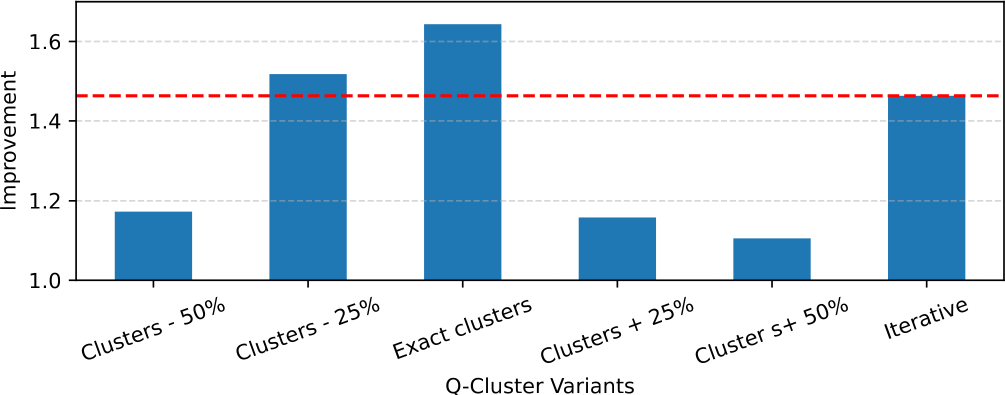}
    \caption{Performance of Q-Cluster compared to Q-Cluster with the number of clusters provided a priori.}
    \label{fig:real_clus}
\end{figure}

Fig.~\ref{fig:raw} shows the fidelity improvement from different QEM schemes alongside Q-Cluster and its variants on low-entropy benchmarks while Fig.~\ref{fig:overall} compares the mean fidelity improvement from different QEM schemes across five IBM devices \textit{ibm\_torino}, \textit{ibm\_brisbane}, \textit{ibm\_strasbourg}, \textit{ibm\_brussels}, and \textit{ibm\_kyiv}, with \textit{ibm\_torino} being from the Heron r1 architecture and other machines from the Eagle r3 architecture. 

The figures demonstrate that Q-Cluster has the best overall performance across all five devices. DD and Pauli Twirling are enabled for all runs across all four machines. 
A quick note is that, due to the long post-processing times of QBEEP, we were not able to finish processing the results for benchmarks with more than 15 qubits. Across different devices, the overall performance of Q-Cluster is consistently better than other state-of-art methods.  
Q-Cluster performs well on low-entropy benchmarks due to its ability to accurately identify both the correct centroids and cluster count. This results in significant fidelity improvements for benchmarks like \textit{bv\_n14}, \textit{ghz\_state\_n23}, \textit{cat\_state\_n22}, and \textit{adder\_n10} across all machines (Fig. \ref{fig:raw}a).

Raw fidelity results in Fig. \ref{fig:raw}b show that Q-Cluster outperforms SOTA methods, particularly when the fidelity and entropy of the noisy distributions are low. This is driven by Q-Cluster's effective clustering and elimination of spurious bit-strings during redistribution.

We also observe that overestimating $p_e$ by 50\% improves performance over the iterative Q-Cluster in nearly all cases, showing the robustness of Q-Cluster on $p_e$ estimation for real-device data.

Lastly, we evaluate Q-Cluster’s performance when the number of clusters is provided a priori, as shown in Fig. \ref{fig:real_clus}. Supplying the exact number of clusters results in a 1.12× improvement over the iterative Q-Cluster, further improving the average fidelity from 0.583 to 0.703. Notably, even providing an estimate that underestimates the ideal number of clusters by up to 25\% still yields improved performance. This confirms that user-provided cluster information can further improve the performance of Q-Cluster. 

\subsection{Runtime of QEM Methods} \label{sec:runtime_res}

Table \ref{tab:comparison_time} presents the post-processing time (in seconds) of different QEM approaches. We ran the post-processing on an Intel(R) Xeon(R) Silver 4216 CPU with 32 logical cores and 64 GB of RAM using the entire dataset and reported the average time for all the low-entropy benchmarks below 15 qubits. The table shows that M3 outperforms the other methods by a significant margin, followed by Q-Cluster, whereas HAMMER and QBEEP are significantly slower. This result aligns with the complexity analysis in Table \ref{tab:1}. QBEEP runs slower because we used the vanilla implementation \cite{QBeep-code}, which pointed out that it could be further optimized. In contrast, the M3 package uses a fully optimized implementation, while HAMMER and Q-Cluster remain unparallelized. Q-Cluster is significantly faster since the number of clusters is always significantly smaller than the number of shots for all the low-entropy benchmarks. 
\begin{table}[ht!]
\centering
\caption{Comparison of QEM methods' Average Execution Time}
\begin{tabular}{|l|c|c|c|c|}
\hline
\textbf{}          & \textbf{M3} & \textbf{HAMMER} & \textbf{Q-Cluster} & \textbf{QBEEP} \\ \hline
\textbf{Time (s)}  & 0.2445         & 15.9342            & 0.8330              & 15.6335         \\ \hline
\end{tabular}
\label{tab:comparison_time}
\end{table}




\section{Conclusions}\label{sec:Conclusion}

In this paper, we propose a novel QEM method based on unsupervised learning to mitigate noisy output distributions from NISQ machines. We show that unsupervised learning, clustering in particular, can identify salient features to recover data patterns from noisy distributions. Our proposed scheme, Q-Cluster, is particularly effective on quantum circuits with low-entropy output distributions. Compared with state-of-the-art techniques, Q-Cluster outperforms M3 by 1.29x, Hammer by 1.47x, and QBeep by 2.65x. We highlight that incorporating noise-tailoring techniques can significantly enhance the performance of QEM methods. 

This study demonstrates the importance of tailoring noise and QEM methods to address specific types of noise, rather than attempting to design algorithms that handle all types of errors.
\section{Acknowledgements}
The work is funded in part by NSF grants 1818914,
2325080 (with a subcontract to NC State University from Duke University), 2120757 (with a subcontract to NC State University from the University of Maryland) and by the U.S. Department of Energy, Advanced Scientific Computing Research, under contract number DE-SC0025384.
\bibliographystyle{IEEEtranS}
\bibliography{references}
\end{document}